\documentclass[runningheads]{llncs}

\usepackage{mdwmath}
\usepackage{eqparbox}

\usepackage{fixltx2e}
\usepackage{stfloats}

\usepackage[pdftex]{graphicx}
\graphicspath{{./images/}}
\DeclareGraphicsExtensions{.jpg,.png}

\usepackage[caption=false,font=footnotesize]{subfig}

\usepackage{amsmath}
\usepackage{bbm}
\usepackage{stmaryrd}

\usepackage[usenames, dvipsnames]{color} 
\usepackage[hyphens]{url}
\usepackage[pdftitle=Title,pdfauthor=Anonymous]{hyperref}

\usepackage{multirow}
\usepackage{tabularx} 
\usepackage{rotating} 

\newcounter{mireya}

\newcounter{ben}

\newcounter{al}

\def\Snospace~{\S{}}

\newcommand{\mAC}[1]{
  \noalign{\vskip 2mm}
  \multicolumn{4}{l}{\textbf{Attack Class} {#1}}
}

\newcommand{\mG}[1]{
  \multicolumn{4}{l}{\textit{Goals:} {#1}}\vspace{0.2 cm}
}

\newcommand{\mTwo}[2]{
  \begin{tabular*}{.4\textwidth}[t]{l}
    {#1} \\
    {#2}
  \end{tabular*}
}
\newcommand{\mThree}[3]{
  \begin{tabular*}{.4\textwidth}[t]{l}
    {#1} \\
    {#2} \\
    {#3}
  \end{tabular*}
}

\begin{document}
\title{The Looming Threat of China: \\ An Analysis of Chinese Influence on Bitcoin}
\author{}
\author{Ben Kaiser\inst{1} \and
	Mireya Jurado\inst{2} \and
	Alex Ledger}
\authorrunning{B. Kaiser et al.}
%
\institute{Princeton University, Princeton, NJ 08544, USA \and
	Florida International University, Miami, FL 33199, USA}
\maketitle


\begin{abstract}
As Bitcoin's popularity has grown over the decade since its creation, it has become an increasingly attractive target for adversaries of all kinds. One of the most powerful potential adversaries is the country of China, which has expressed adversarial positions regarding the cryptocurrency and demonstrated powerful capabilities to influence it. 
In this paper, we explore how China threatens the security, stability, and viability of Bitcoin through its dominant position in the Bitcoin ecosystem, political and economic control over domestic activity, and control over its domestic Internet infrastructure. We explore the relationship between China and Bitcoin, document China's motivation to undermine Bitcoin, and present a case study to demonstrate the strong influence that China has over Bitcoin. Finally, we systematize the class of attacks that China can deploy against Bitcoin to better understand the threat China poses. We conclude that China has mature capabilities and strong motives for performing a variety of attacks against Bitcoin.
\end{abstract}

\thispagestyle{plain}
\pagestyle{plain}



\section{Introduction}

%

In 2008 Satoshi Nakamoto published the Bitcoin white paper, using cryptography to create the world's first decentralized currency~\cite{Nakamoto09}.
Since its creation, Bitcoin's popularity has grown substantially, reaching a market capitalization of over \$100 billion USD as it continues to attract interest from technology enthusiasts, black markets, and legitimate markets~\cite{blocksci-market-cap}.

Despite this popularity, the security of Bitcoin is still not fully understood. Many serious attacks have been theorized by researchers but have not yet come to pass, leading to the adage ``Bitcoin is secure in practice but not in theory.'' Some analyses approach this strange characterization of Bitcoin by saying that the security model of Bitcoin must consider the socioeconomic and political forces in addition to the underlying cryptography~\cite{Bonneau15,bonneau2018hostile}.

The decentralized nature of Bitcoin presents unique socioeconomic and political challenges. Operation and maintenance tasks are distributed across a massive number of peers called miners, and because there is no central governing structure, these miners are kept honest by a carefully balanced incentive scheme. The system is designed so that anyone can contribute by devoting some computing power to mining, but over the last several years, Bitcoin mining has become heavily centralized due to advances in specialized hardware that render commodity hardware obsolete. As a result, miners have congregated into mining pools: consortia of miners who work together and share profits.
As of June 2018, over 80\% of Bitcoin mining is performed by six mining pools~\cite{blocksci-hashrate-distribution}, and five of those six pools are managed by individuals or organizations located in China.

One broadly understood security property of Bitcoin is that no single party can control more than 50\% of the hash rate, so this statistic is worrying. The Chinese government exerts strong, centralized control over economic and financial activity and also operates extensive surveillance and censorship regimes over the domestic Internet. These capabilities do not grant them direct command of all of the hash power in Chinese-managed pools, but they do have a variety of tools at their disposal to influence those pools and Bitcoin in general. They have deployed multiple rounds of restrictive regulations that have upended global and domestic Bitcoin markets, and as we show in \autoref{sec:technical-interference}, Chinese Internet surveillance has affected transaction throughput.

This interest and activity in Bitcoin by China raises many questions. Where does China actually stand on Bitcoin? Do they have motivations to influence Bitcoin globally, and have they succeeded in doing so in the past? What capabilities do they have to influence Bitcoin?

\textbf{Contributions:} In this paper, we explore whether and how China threatens the security, stability, and viability of Bitcoin through its position in the Bitcoin ecosystem, political and economic control over domestic activity, and technical control over its domestic Internet infrastructure.
We find that China has motivations to threaten Bitcoin and has influenced it through domestic regulatory and technical measures. We also show that China has a number of mature capabilities for executing a variety of attacks. We enumerate and classify the attacks that China can execute and what goals that they would achieve. Finally, we outline directions for future research, including additional dimensions of analysis and surveying potential mitigations to the threat China poses.



\textbf{Organization:}
The paper is organized as follows. \autoref{sec:bitcoin} summarizes the design and operation of Bitcoin. \autoref{sec:china} briefly summarizes China's relationship with Bitcoin and the technical and non-technical capabilities China could bring to bear on Bitcoin. \autoref{sec:technical-interference} presents an analysis of how China's Internet censorship limited throughput for Bitcoin as a whole, demonstrating a link between Chinese domestic policy and the global stability of Bitcoin. \autoref{sec:risks} systematizes the risk that China poses to the Bitcoin ecosystem. Finally, \autoref{sec:conclusion} concludes and outlines directions for future work.

\section{Bitcoin}
\label{sec:bitcoin}

Bitcoin is a distributed ledger that tracks payments in a digital currency.
Whereas traditional currencies and payment systems rely on monolithic financial institutions to control supply and mediate transactions, Bitcoin distributes those responsibilities among a set of peers called miners who are rewarded for their efforts by receiving payouts in the currency.
This approach
diffuses trust, allowing users to place small amounts of trust in many different parties rather than all of their trust in a single entity.

\subsection{Technical overview}
\label{subsec:bitcoin-technical-overview}

As outlined in \cite{Bonneau15}, Bitcoin consists of three components: transactions transferring ownership of coins, the consensus protocol, and the communications network.

\begin{raggedleft}
\textbf{Transactions:}
Bitcoin transactions are protocol messages that transfer currency called bitcoins (abbreviated BTC) from one user to another.
Each user is represented by a public/private key pair, and the hash of the public key serves as an address that can be associated with transaction inputs and outputs.
By creating a new transaction, a user can claim bitcoins output from a prior transaction if they prove ownership of the corresponding private key.
\end{raggedleft}

\begin{raggedleft}
\textbf{Consensus protocol:}
In order to maintain consensus, transactions are organized into a shared public ledger, implemented as a series of blocks.
Each block contains a set of transactions, a timestamp, an arbitrary number called a nonce, a hash of the previous block, and some other protocol information.
Storing the previous hash means that each block is cryptographically linked to its predecessor, creating the blockchain data structure.
Integrity is preserved through the blockchain since any alteration to an earlier block will cause its hash to change and therefore `break' the chain.
\end{raggedleft}

Miners collect transactions, check their validity, organize them into blocks and publish them to the network.
They are rewarded for the effort of creating blocks by a special transaction in each block that sends a quantity of bitcoins to an address of the miner's choosing.
This is called the block reward, and it is the only means by which new bitcoins are created.

To determine which block is next in the chain to maintain global consensus, Bitcoin uses a computational puzzle called \textit{Proof of Work}.
A valid block is defined as one whose double SHA256 hash is below a target threshold value.
Miners fix all of the fields of the block except for the nonce, then randomly guess different nonces until a valid block is found.
If multiple valid next blocks are mined, a fork occurs and other miners must choose which branch to mine off of.
To resolve forks, miners simply mine from the longest branch.

Typically miners collaborate in groups called mining pools.
Pool members submit partial proofs-of-work (PPoWs), which are blocks that hash to a value close to the target but are not actually valid.
PPoWs serve to measure the amount of work that a miner has been conducting in the effort to find a block.
They are sent to a pool manager who allocates rewards to members in proportion to the computational work they performed.

\begin{raggedleft}
\textbf{Communication network:}
Bitcoin nodes use a peer-to-peer broadcast network to announce and propagate transactions and blocks.
Nodes in the communications network follow a set of rules to enhance performance and support the consensus protocol; for example, they will only forward new data once to prevent infinite propagation and will only relay valid data to prevent invalid blocks or transactions from being spread on the network.
\end{raggedleft}

The communication network's performance and degree of centralization have an effect on the consensus protocol.
With respect to performance, high latency between nodes can cause temporary forks which in turn cause instability.
With respect to centralization, centralized control of nodes or the connections between them can affect the fairness of the protocol.
For example, if a miner controls enough of the nodes, they can favor their own blocks to win the forks and earn the block reward.
Similarly, if anyone is able to censor the network, they can prevent blocks and transactions from spreading.
Thus, to ensure stability and fairness, Bitcoin requires a low latency, decentralized, uncensorable network.

\begin{raggedleft}
\textbf{Usage of Bitcoin:}
In practice, some people use Bitcoin as a store of value, but it is mostly used as a speculative investment asset.
It remains largely divorced from existing monetary systems, although in some places it has gained a foothold.
In countries experiencing high degrees of inflation, Bitcoin and other cryptocurrencies gain popularity because they do not rely on the country's financial infrastructure and their value cannot be manipulated by the government.
In Venezuela, for example, some citizens have converted rapidly inflating Bolivars into Bitcoin to store value or purchase goods online, and mining has become increasingly common as a source of income~\cite{Torpey18}.
Similar trends have been observed when similar economic conditions have occurred in Greece, Zimbabwe, and Ukraine~\cite{Armario17}.
In the US, investment banks have begun to tentatively embrace Bitcoin, with Goldman Sachs leading the way by first facilitating Bitcoin trading on other platforms and soon offering its own Bitcoin derivative products to clients~\cite{Popper18}.
\end{raggedleft}


\section{Bitcoin in China}
\label{sec:china}
\begin{figure}[t]
	\label{fig:china}
	\centering
	\subfloat[][]{
		\includegraphics[width=0.49\textwidth]{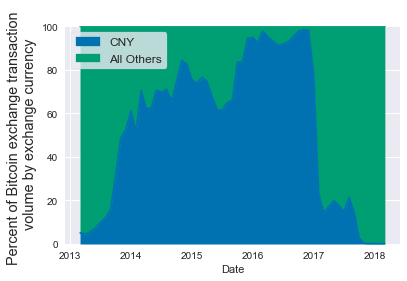}
		\label{fig:chinaexchange}
	}
	\subfloat[][]{
		\includegraphics[width=0.49\textwidth]{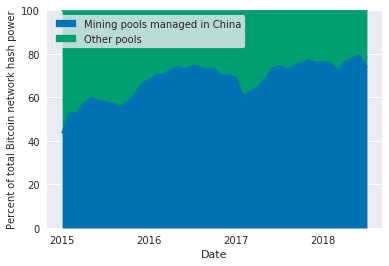}
		\label{fig:hashpower}
	}
    \caption[]{
   		(a) Percentage of global bitcoin exchange volume processed in CNY from 2013-2018, using data from \cite{bitcoinityExchange}.
   		(b) Percentage of total Bitcoin hash power controlled by pools managed within China\protect\footnotemark, using data from \cite{BTChashrate}.
    }
\end{figure}

Bitcoin's rise in China began in 2013. In the following years, Chinese exchanges grew to dominate the global exchange market, as shown by the relative share of Bitcoin exchange transactions executed in Chinese Yuan (CNY) versus other currencies (Figure~\ref{fig:chinaexchange}). Mining pools managed by individuals in China have constituted over half of the total network hash power since 2015 (Figure~\ref{fig:hashpower}) and currently more hash power is located in China than in any other country~\cite{Hileman17}.

Through this time, China's official position on Bitcoin remained ambiguous and regulators proved unwilling to institute tight controls despite expressing concerns over criminal activity, subversion of capital controls, and speculative risk. This tenuous equilibrium between demand by Chinese users and investors and intermittent regulatory impedance shaped Bitcoin's global trajectory until it was punctured in 2017 by firm regulations on the exchange industry. Appendix 1 provides a more detailed discussion of the Bitcoin exchange and mining sectors in China over this period. In this section, we identify how China's dominant position in the Bitcoin ecosystem  and tight control over domestic economic and technical resources grant them capabilities to influence Bitcoin.


\footnotetext{We attribute the following pools to Chinese managers: AntPool, Bixin, BTCC, BTC.com, BTC.TOP, BW.COM, DPOOL, F2Pool, Poolin, ViaBTC, and 58COIN.}

\begin{flushleft}
	\textbf{Regulatory authority:}
	The Chinese government enjoys broad regulatory authority that it can bring to bear on domestic Bitcoin users, exchanges, and miners. Regulators have issued policy decrees to directly influence the exchange and mining sectors and also targeted Bitcoin indirectly through externalities like energy prices (see Appendix 1 for details).
\end{flushleft}

\begin{flushleft}
	\textbf{Internet traffic tampering and surveillance:}
	China operates a variety of Internet control measures that can affect Bitcoin traffic. The most well-understood system is the Great Firewall (GFW), which performs on-path surveillance and traffic filtering using deep packet inspection (DPI) and active probing of connection endpoints~\cite{ensafi2015analyzing}. As an on-path tool, the GFW can observe network traffic and inject new packets but it cannot prevent packets that have already been sent from reaching their destination. For more active traffic tampering, China operates a separate in-path tool known as the Great Cannon, which can inject malicious code into packets in transit and levy denial-of-service attacks by redirecting traffic to a target host~\cite{Marczak15}. Both of these systems primarily operate on traffic transiting between China and the rest of the world, but central government regulators also control all Internet Service Providers (ISPs) in China, allowing for collection and analysis of domestic traffic.
\end{flushleft}







\begin{flushleft}
	\textbf{Hash power:}
	At the time of writing, 74\% of the hash power on the Bitcoin network is in Chinese-managed mining pools. Pool miners cannot be directly controlled by China, but the managers are located within China and as such are subject to Chinese authorities. Because managers are responsible for assigning mining jobs and propagating completed blocks, they control the inputs and outputs of their miners, allowing Chinese authorities indirect control over that hash power.
	China has more direct control over the hash power physically located in China. This is a significant share of the global hash rate -- more than controlled by any other single country~\cite{Hileman17} -- but the precise quantity is unknown.
\end{flushleft}

\begin{flushleft}
	\textbf{Well-connectedness within the Bitcoin network:}
	Which blocks reach consensus in Bitcoin depends in part on how quickly they propagate through the network from their source miner to other peers. Blocks found in China are already proximate to a majority share of hash power, so they can reach consensus more quickly than blocks found elsewhere. If the Chinese government assumed control of domestic hash power, this property would grant them an advantage in selecting blocks for the ledger, which is important for some types of attacks (see~\autoref{sec:risks}).
\end{flushleft}



\section{Technical Interference}
\label{sec:technical-interference}
In this section, we demonstrate a case of Chinese technical interference in Bitcoin. Specifically, we discuss how the GFW imparts a latency overhead on all traffic it processes, including Bitcoin traffic. Until protocol upgrades were introduced to address this problem in June 2016,
Chinese miners were disadvantaged by this latency as it slowed the rate at which their proposed blocks could propagate. This created an incentive for those miners to mine empty blocks because those were less disadvantaged by the latency. However, empty blocks are bad for Bitcoin, as they process no transactions but consume network resources, thus damaging system throughput.

\subsection{Analysis}
\begin{figure}[t]
	\centering
	\subfloat[][]{
		\includegraphics[width=0.49\textwidth]{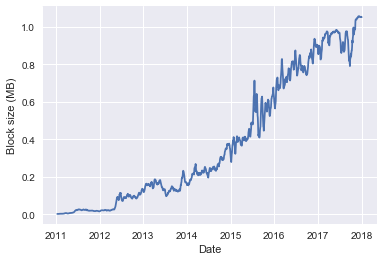}
		\label{fig:avgblocksize}
	}
	\subfloat[][]{
		\includegraphics[width=0.49\textwidth]{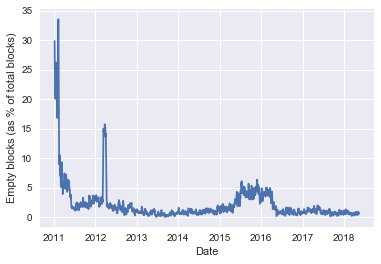}
		\label{fig:pctemptyblocks}
	}
	\caption{
		(a) Average block size in MB, averaged weekly, using data from \cite{BCInfo-BlockSize18}.
		(b) Percentage of published blocks per day that are empty, using data from \cite{Blockchair18}.
	}
	\end{figure}

Bitcoin blocks adhere to a fixed size limit of 1MB.
\footnote{The SegWit protocol upgrade, activated in August 2017, kept the 1MB limit for transactions but allowed other block data to consume an additional 3MB.
This is why Figure~\ref{fig:avgblocksize} shows that some recent blocks are larger than 1MB.}
Since mid-2016, blocks on the main chain have generally been at or above 800KB in size, suggesting that throughput is at 80\% of capacity, although prior to this smaller blocks were the norm (see Figure~\ref{fig:avgblocksize}).

Miners are free to include as many or as few transactions in a block as they want. Because the difficulty of mining a block does not depend on its size, to maximize their profits miners would be expected to include as many fee-bearing transactions in their blocks as possible (even though transaction fees are quite small compared to the block reward). If there are not enough fee-bearing transactions in the mempool to fill a block, a rational miner might publish a block with some empty space because although including additional non-fee-bearing transactions does not cost them anything, it also does not earn them anything.

Based on this incentive structure, empty blocks (those containing only the coinbase transaction) should be exceedingly rare, as there are virtually always some fee-bearing transactions in the mempool. However, historically empty blocks were occasionally produced. There are reasons a miner might do this and forgo transaction fees; for example, once a miner is notified that a block has been found by another miner, they may start mining an empty block in the time it takes to download and validate that block and choose the next set of transactions. Although this process is typically fairly quick, the probabilistic nature of mining means that sometimes the miner successfully finds that empty block in time, and so they publish it to reap the block reward.

Prior to May 2015, empty blocks were produced at a consistent rate of $2 - 3\%$, except for a few aberrations in 2011 and 2012 before Bitcoin was widely used (see Figure~\ref{fig:pctemptyblocks}). Beginning in May 2015, a noticeable spike up to around $5\%$ that lasts through June 2016 is visible.
We examined the behavior of the eight largest mining pools during this period, which cumulatively found nearly 80\% of all blocks. Table~\ref{tbl:hashrate15-16} summarizes this information. The four Chinese pools cumulatively account for 64\% of the hash power while non-Chinese pools account for 24\%.

\begin{figure}[t]
	\centering
	\subfloat[][]{
		\includegraphics[width=0.49\textwidth]{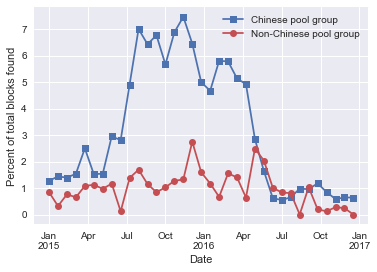}
	 	\label{fig:pctemptyblocksgrouped15-16}
	}
	\subfloat[][]{
		\includegraphics[width=0.49\textwidth]{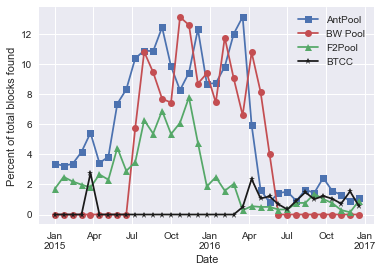}
		\label{fig:pctemptyblockschinese15-16}
	}
	\caption{
	(a) Percentage of published blocks that are empty, sampled once per three weeks and grouped by national affiliation, using data from
	\cite{Blockchair18}. See Table \ref{tbl:hashrate15-16} for information on mining pool hash rates and national affiliations.
	(b) Percentage of blocks published by Chinese-controlled pools that are empty, sampled once per three weeks, using data from \cite{Blockchair18}.
	}
\end{figure}

Looking at the combined average rates of empty blocks produced by each of these pool groups (Figure~\ref{fig:pctemptyblocksgrouped15-16}), we see that the Chinese pool group produced an unusually high rate of empty blocks, spiking up above 7\%.
\footnote{Within the group, two individual pools (AntPool and BW Pool) peaked with empty block rates as high as 13\% (see Figure~\ref{fig:pctemptyblockschinese15-16}).}
Meanwhile, non-Chinese miners produced empty blocks at a historically consistent rate of around 2\%.
These observations suggest that some factor that applied to Chinese miners -- but not other miners -- created an incentive to mine empty blocks. We posit that this factor is the Great Firewall, and more specifically, the bandwidth bottleneck it imparts.

\begin{table}[t]
	\centering
	\caption{Nationality of mining pools and their share of hash rate between May 1, 2015 and June 30, 2016.}
	\label{tbl:hashrate15-16}
	\begin{tabular}{|l|l|l|}
		\hline
		\textbf{Mining pool} & \textbf{Located in China} & \textbf{\begin{tabular}[c]{@{}l@{}}Estimated share of\\ network hash rate\end{tabular}} \\ \hline
		F2Pool               & Yes                       & 22.17                                                                                   \\ \hline
		AntPool              & Yes                       & 21.54                                                                                   \\ \hline
		BTCC                 & Yes                       & 12.79                                                                                   \\ \hline
		BitFury              & No                        & 12.39                                                                                   \\ \hline
		BW Pool               & Yes                       & 7.84                                                                                    \\ \hline
		KnCMiner             & No                        & 4.89                                                                                    \\ \hline
		SlushPool            & No                        & 4.72                                                                                    \\ \hline
		21 Inc.              & No                        & 2.27                                                                                    \\ \hline
	\end{tabular}
\end{table}

The GFW is known to limit bandwidth by inducing packet loss in TCP streams. In 2017 a test showed 6.9\% packet loss in connections between the US and China and only 0.2\% for connections between the US and Hong Kong, which is just outside the GFW \cite{Xu17}. Packet loss causes latency, as dropped packets must be re-requested and re-sent, and this was also observed in the same test (218ms latency for Chinese agents compared to 81ms for Hong Kong).
This latency has been observed to affect Bitcoin block propagation. In 2015 it was shown that block propagation across the GFW was an order of magnitude slower than propagation between nodes on the same side of the GFW \cite{ToomimScalingBTC2015}. Separate research the following year found that mean propagation times for near-full Bitcoin blocks were 3.9 seconds between nodes on the same side of the GFW and 17.4 seconds between nodes on opposite sides, representing a slowdown of nearly 450\% \cite{Rizun2016}.

Propagation latency perverts the incentives for miners behind the GFW because they are at an inherent disadvantage. If a miner behind the GFW finds a valid next block at the same time as an outside miner, the latter can more quickly propagate their block to the rest of the network, meaning they are more likely to win a forking race and have their block accepted.
Mining smaller blocks, which propagate more quickly, can counteract this disadvantage for miners behind the GFW. While transaction fees are designed to dissuade empty block mining, those fees fluctuate and at times are insubstantial. Across the period when empty block rates were high, transaction fees averaged only .25 BTC per block. The block reward of 25 BTC was far more valuable, so mining more empty blocks to earn more block rewards would seem profitable.

Compact block relay was incorporated into the Bitcoin Core client on June 22, 2016 as BIP152 (Bitcoin Improvement Propsal \#152) \cite{BIP152PR}. Leveraging the fact that full nodes store transactions they have already seen in their mempool, the BIP significantly reduced the bandwidth required for block propagation. Specifically, it permitted peers to avoid sending full blocks to each other and instead send compact block ``sketches'' along with a small set of transactions the sender guesses the receiver has not yet seen. Under this protocol, a full block can be relayed using only 15KB of data \cite{Gomez16}, reducing the bandwidth requirement of block propagation by 98\%.

With BIP152 incorporated, block propagation time became largely independent of block size, and so the incentive to mine empty blocks was eliminated. Correspondingly, the practice quickly subsided to its pre-2015 level (as seen in Figures~\ref{fig:pctemptyblocksgrouped15-16} and~\ref{fig:pctemptyblockschinese15-16}).
While this particular threat to Bitcoin has abated, it demonstrates the power of China's technical capabilities for domestic control to weaken Bitcoin, even unintentionally, on a global scale. We now turn to an analysis of other vectors by which China could leverage their capabilities to attack Bitcoin.


\section{Threats}
\label{sec:risks}
There are many known attacks on Bitcoin's consensus, miners, software clients, and communication network (see~\cite{Conti17} for a survey).
In this section, we catalog four classes of attacks that China could perpetrate on Bitcoin using the capabilities outlined in \autoref{sec:china} and posit a set of metrics for understanding the attack classes.


\subsection{Metrics}
\textbf{Goals:} There are four goals that China may wish to achieve by attacking Bitcoin. First, as discussed in Appendix 1, Bitcoin stands in ideological opposition to China's centralized governing philosophy, so they may be motivated to weaken or destroy it to make an \textit{ideological statement}; for example, demonstrating the futility of decentralized control paradigms. Virtually any violation of Bitcoin's security suffices to achieve this goal as long as it is highly visible.

Second, the government may attack Bitcoin for the purpose of \textit{law enforcement}: administering capital controls or preventing other illegal activity. Targeting specific users for deanonymization and censorship would allow China to crack down on illicit uses of Bitcoin.

Third, although China has expressed distrust of Bitcoin, they may still determine that \textit{increasing control} over the system is beneficial (e.g., to achieve other goals outlined in this section). By disrupting non-Chinese mining pools, especially those with significant hash power, China could further increase the proportion of hash power they can control and thus exert more influence over Bitcoin.

Finally, as Bitcoin becomes more widely used and more tightly integrated into global financial systems, it becomes a possible vector for attacking foreign economies. To \textit{exert influence in a foreign country} where Bitcoin is in use, China may aim to weaken or even totally destroy Bitcoin. This could be done by targeting specific users or miners for attack or by generally weakening consensus to increase volatility to a breaking point.

\begin{raggedleft}
\textbf{Visibility:}
Some attacks on Bitcoin can be performed surreptitiously while others are easily detectable.
We distinguish \textit{covert} attacks, which are difficult to detect and only minimally perturb the Bitcoin network or ledger, from \textit{overt} attacks which produce visible signatures suggesting Chinese culpability.
Note that any attack can be overt if China announces their actions; we do not classify an attack as overt if this is the only way it can be linked to China.
\end{raggedleft}

\begin{raggedleft}
\textbf{Targets:}
An attack can target Bitcoin users, miners, or the entire ecosystem.
\end{raggedleft}

\begin{raggedleft}
\textbf{Capabilities:}
We discuss China's capabilities to effect Bitcoin in \autoref{sec:china} and note the difficulty of accurately estimating hash power under their control. For our attack analysis, we divide attacks dependent on hash rate into three thresholded categories.
If an attack requires the majority of the network hash rate (i.e., 51\% or more), we label that as \textit{high threshold}.
Attacks that are more effective with the majority of the hash rate but are possible with less have a \textit{medium threshold}, and attacks requiring significantly less than a majority of the hash rate are \textit{low threshold}.
\end{raggedleft}

\subsection{Attacks}
\label{subsec:attacks}

  \begin{table}
    \caption{A taxonomy of attacks China can deploy to influence the Bitcoin ecosystem.}
    \label{tbl:attacks}
    \centering
    \begin{tabular*}{\textwidth}{ l  l  l  l }
      \hline
      \mAC{1. Censor specific users or miners} \\
      \mG{Ideological statement, law enforcement, foreign influence} \\
      \hline
      \textbf{Attacks} & \textbf{Visbility} & \textbf{Target} & \textbf{Capabilities} \\
      (a) Punitive forking & Overt & Users & High hash rate \\
      (b) Feather forking & Overt & Users & Low hash rate \\
      (c) Eclipse attack & Covert & Users & \mTwo{Control over a large}{number of peers} \\
      \mTwo{(d) Internet traffic}{ tampering} & Overt & Users & \mTwo{Internet traffic}{tampering} \\
      \hline
      \mAC{2. Deanonymize users}\\
      \mG{Ideological statement, law enforcement}\\
      \hline
      \textbf{Attacks} & \textbf{Visbility} & \textbf{Target} & \textbf{Capabilities} \\
      (a) Heuristic address clustering & Covert & Users &
       Compute power  \\
      (b) Traffic monitoring & Covert & Users & Internet surveillance \\
      \mTwo{(c) Compel service providers}{ to deanonymize customers} & Covert & Users & Coercion/regulation \\
      \mTwo{(d) Third-party tracking}{ of Web purchases} & Covert/Overt  & Users & \mThree{Internet surveillance;}{Coercion/regulation;}{Tracker injection} \\
      \mTwo{(e) Compel users directly}{to deanonymize} & Covert & Users & Coercion/regulation \\
      \hline
      \mAC{3. Weaken consensus / Destabilize Bitcoin} \\
      \mG{Ideological statement, foreign influence} \\
      \hline
      \textbf{Attacks} & \textbf{Visbility} & \textbf{Target} & \textbf{Capabilities} \\
      (a) Race attack & Overt & Users & \mTwo{Medium hash rate;}{Well-connectedness} \\
      (b) Finney attack & Overt & Users & \mTwo{Medium hash rate;}{Well-connectedness} \\
      (c) Brute force attack & Overt & Users & \mTwo{Medium hash rate;}{Well-connectedness} \\
      (d) Balance attack & Overt & Users & \mThree{Low hash rate;}{Internet traffic}{tampering} \\
      (e) Goldfinger attack & Overt & Ecosystem \hspace{1mm} &  High hash rate\\
      (f) Selfish mining & Overt & Miners & \mTwo{Low hash rate;}{Well-connectedness} \\
      (g) Eclipse attack & Covert & Miners & \mTwo{Control over a large}{number of peers}  \\
      \hline
      \mAC{4. Disrupt competing mining operations} \\
      \mG{Increase control} \\
      \hline
      \textbf{Attacks} & \textbf{Visbility} & \textbf{Target} & \textbf{Capabilities} \\
      (a) Selfish mining & Overt & Miners & \mTwo{Low hash rate;}{Well-connectedness} \\
      (b) Block withholding & Covert/Overt \hspace{1mm} & Miners & Low hash rate\\
      (c) Fork after withholding & Covert/Overt & Miners & Low hash rate \\
    \end{tabular*}
  \end{table}

\autoref{tbl:attacks} summarizes the different classes of attacks that are feasible for China, specifies the technical means by which each could be achieved, and connects them to the goals outlined above. Further discussion of each individual attack class follows.

\begin{raggedleft}
\textbf{Censorship:}
One class of attacks in China's arsenal is the ability to perform targeted censorship of Bitcoin users, preventing them from committing transactions to the blockchain. By censoring specific users, China could achieve three possible goals: first, they could make an ideological statement that even decentralized ecosystems like Bitcoin are still subject to China's centralized control; second, they could censor Bitcoin addresses known to belong to criminals to crack down on illegal activity; and third, they could weaken organizations or foreign economies that rely on Bitcoin by selectively censoring addresses important to those parties.
\end{raggedleft}

With control of at least 51\% of the hash rate, Chinese mining pools could simply announce that they will not mine on chains containing transactions from their list of censored addresses. This is called a \textit{(a)} \textit{punitive forking} attack. With less than 51\% of the hash power, Chinese miners could still attempt to fork whenever they see a censored transaction, but some attempts may fail. However, the forks that succeed orphan the blocks found by miners that include censored transactions, reducing their profits, so some may be convinced to follow China's censorship rules. This is a \textit{(b)} \textit{feather forking} attack~\cite{millerforum,narayanan2016bitcoin}. As both attacks require announcing intent, we classify them as overt.

One way China could reduce the hash power required for forking-based censorship attacks is through an \textit{(c)} \textit{eclipse attack}~\cite{Heilman15}.
By directing a large number of peers to monopolize all incoming and outgoing connections to specific victim nodes, this attack controls what those victim nodes see and do in the Bitcoin network in order to prevent them from learning about the transactions China wants to censor.
This reduces the portion of the network that is counteracting censorship attacks by trying to approve the transactions China is trying to censor, meaning that less hash power is required to succeed at censorship, especially if the targeted victim nodes are miners with substantial hash power.
This attack can be performed covertly, as victimized peers are unlikely to realize that their connections are being manipulated.

The final attack that China can employ for censorship is \textit{(d)} Internet traffic tampering using the GFW and control over domestic ISPs. China could either block blacklisted transactions originating in China from propagating or prevent blacklisted transactions originating outside of China from entering the country and reaching Chinese miners. This attack is overt because it would be clear to Bitcoin users if Chinese miners were not adding their transactions to their blocks.

\begin{raggedleft}
\textbf{Deanonymization:}
Bitcoin is designed to preserve the pseudonymity of its users, meaning that their real-world identity cannot be linked to a Bitcoin address they have used to transact.
However, in practice there are complications that make deanonymization attacks possible.
\end{raggedleft}

China might seek to deanonymize users for two reasons. First, they may wish to enforce laws and regulations; for example, enforcing capital flight restrictions by identifying users purchasing foreign goods or exchanging Bitcoin into foreign currencies. They might also use a deanonymization attack for ideological (or political) ends: to publicly reveal malfeasance by subversives or political opponents or simply to demonstrate the superiority of centralized control as an ideology and discourage enthusiasm for decentralized systems.

We identify four attacks that China could use to deanonymize specific users.
First, they could use known research techniques to \textit{(a)} heuristically cluster pseudonymous identities (e.g., connect multiple addresses to the same user)~\cite{Meiklejohn13,Ron13}.
The simplest example of such a heuristic is to cluster addresses that appear as multiple inputs to the same transaction, as they presumably belong to the same user.
The only required capabilities are access to the blockchain and marginal compute power to run the analytics, so these attacks are not unique to China; virtually anyone could commit them.

Where China has an advantage over typical adversaries is in linking these pseudonyms to IP addresses. One approach would be to covertly \textit{(b)} monitor Bitcoin network traffic and identify which IP addresses transactions originate from~\cite{Biryukov14,Koshy14}.
Because Bitcoin traffic is unencrypted, this can be done through deep packet inspection (DPI).
China could also use \textit{(c)} coercion or regulation to covertly compel service providers that deal in Bitcoin, such as merchants or exchanges, to identify their users. 
Further, it has been shown that when Bitcoin is used for online purchases, enough information is leaked to web trackers that they can uniquely identify the transaction on the blockchain and link it to any identifying information provided by the purchaser~\cite{Goldfeder17}. China could covertly \textit{(d)} intercept this tracking information over the Internet (using DPI) to perform the same attack, compel domestic tracking companies to provide the information (also covertly), or inject their own trackers into Internet traffic to collect similar information themselves. Tracker injection could be detected by anyone specifically monitoring Internet traffic for such attacks, so we note that it would be overt.

Finally, China could target users directly using \textit{(e)} coercion or regulation to compel them to deanonymize themselves or their transaction partners. Again, as long as targets are compelled to keep quiet about orders to reveal information, this attack is covert.

\begin{raggedleft}
\textbf{Undermine consensus and destabilize Bitcoin:}
Nakamoto consensus maintains a consistent and irreversible ordering of approved transactions, which is essential for Bitcoin to be usable as a means of transacting and for distributing mining rewards fairly.
Thus, both users and miners are heavily invested in keeping consensus strong.
\end{raggedleft}

When consensus is weak, there is disagreement about the set or ordering of accepted transactions, which is destabilizing to the whole system.
It causes forks to occur more frequently as nodes with different views of the blockchain resolve their conflicts and introduces the opportunity for a \textit{double-spend attack}, in which an attacker spends the same coins in two different transactions.
A successful double-spend would be catastrophic as users would no longer have confidence that accepted transactions were truly irreversible.
Thus, any weakening of Bitcoin consensus would be destabilizing and could cause the eventual destruction or abandonment of Bitcoin.
This means that the goals China could achieve by such an attack are limited: destabilizing or destroying Bitcoin could only make an ideological statement (demonstrating the futility of decentralized paradigms) or serve as an attack on institutions that rely on Bitcoin.

We begin by discussing the different ways China could execute a double-spend.
Typically, in these attacks the first transaction ($T_1$) is used to make a purchase from a merchant, and once the purchase is fulfilled, the second transaction ($T_2$) sends the coins to an address controlled by the attacker.
If the merchant sees a block containing $T_1$ and immediately confirms it, the attacker can quickly issue $T_2$ and hope that it gets included into a block that miners then continue to mine off of, orphaning $T_1$'s block.
This is a type of double-spend called a \textit{(a)} \textit{race attack}, and to pull it off, China would apply their hash power to find both blocks quickly and their well-connectedness within the Bitcoin network to quickly propagate those blocks.

To mitigate the risk of double-spends, merchants are encouraged to wait for a number of additional blocks to be found before accepting any transaction.
This is called \textit{$n$-confirmation}, where $n$ is the number of blocks that the merchant waits.
Techniques exist to perform double-spends even against n-confirmation merchants.
One example is a \textit{(b)} \textit{Finney attack}, where a miner with significant hash power finds separate blocks containing $T_2$ and $T_1$, broadcasts only the second, then later broadcasts the first after the merchant has completed the purchase. Other miners must pick up on the $T_2$ block for it to win the forking race, but the attacker can also privately mine a chain containing $T_2$ and only publish it after it reaches length $n$ (a \textit{(c)} \textit{brute force attack}). An attacker with close to 51\% hash power can succeed with these attacks probabilistically, increasing their odds if they also have an advantage in quickly propagating blocks as Chinese miners do.

Finding multiple consecutive blocks more quickly than the rest of the network requires a significant share of the hash rate.
Because China also has the ability to tamper with network traffic, they can pull off a double spend with a lower threshold for hash power using a \textit{(d)} \textit{balance attack}~\cite{Natoli16}.
Here, China would disrupt communication between two mining groups, then issue $T_1$ to one group ($G_1$) and $T_2$ to the other group ($G_2$).
Each group would be building a valid chain containing their respective transaction, and after some time China would dedicate their hash power to $G_2$, outpacing $G_1$ and invalidating $T_1$.
Using the GFW to disrupt cross-border Bitcoin traffic would allow China to execute this attack against foreign merchants by setting $G_1$ to be miners outside of the GFW and $G_2$ to be domestic miners.

With a majority share of the hash rate, China could execute a \textit{(e)} \textit{Goldfinger attack} in which they apply their hash power to arbitrarily control the system.
According to the analysis of Kroll et al. in \cite{Kroll13}, Bitcoin can only survive such an attack if the remainder of the miners are willing to pay a cost greater than what China is willing to pay to pull off the attack.
Because other Bitcoin miners are loosely organized and China can bring massive resources to bear, the most likely scenario is a death spiral in which China can credibly threaten a Goldfinger attack and rational miners will be scared off, thus destroying Bitcoin.

Finally, China could use a more subtle attack to weaken Bitcoin consensus: \textit{(f)} \textit{selfish mining} (also known as \textit{block discarding}).
A selfish miner keeps found blocks secret until they find enough consecutive blocks to outpace (or match the length of) the public chain.
They sacrifice revenue to do so, but can reduce others' revenue even more, incentivizing those miners to join the selfish coalition and increase the attacker's power to execute other destabilizing attacks.
\footnote{Eyal et al.~\cite{Eyal13} showed that if 50\% of the network mines on the attacker's chain, the attacker only needs 25\% of the network hash power to succeed.}
This attack is overt as it produces an noticable signature of forks in the blockchain.

All of these attacks can be made easier using the eclipse attack discussed above.
By targeting miners with significant hash power and controlling which transactions they see, China could prevent them from contributing their hash power to forks that China is trying to orphan.
According to Heilman et al.~\cite{Heilman15}, this permits a Goldfinger attack to be achieved with only 40\% of the network hash power, and it similarly lowers the threshold for the other forking-based destabilization attacks we have described.

\begin{raggedleft}
\textbf{Disrupt competing miners:}
In addition to targeting Bitcoin users, China can attack other mining pools in order to consolidate their control over Bitcoin and make other attacks easier.
The \textit{(a)} selfish mining strategy described above is one way to achieve this goal; as miners losing profits join the more profitable Chinese-controlled pools, they also enter China's zone of control.
\cite{Conti17} outlines a number of more direct attacks on competing mining pools, and in this section we highlight two that China is capable of.
\end{raggedleft}


In both attacks, China could direct their hash power to pose as mining participants in other pools and then undermine those pools.
The simplest version is a \textit{(b)} \textit{block withholding} attack, where Chinese miners submit partial proofs-of-work (PPoWs) but do not submit full blocks when they find them \cite{rosenfeld2011analysis}.
This may not arouse suspicion because the probabilistic nature of mining means that it is reasonable that a given miner finds many PPoWs but no full blocks, but the mining pool will be missing out on block rewards. This may cause that pool's miners to abandon mining or switch to a more profitable (possibly Chinese-controlled) pool.

The other possibility is that Chinese miners posing as contributors to a foreign pool could wait to submit a found block until a miner outside the pool broadcasts one, creating a fork.
In the paper describing this \textit{(c)} \textit{fork after withholding} attack, Kwon et al.~\cite{kwon17beselfish} show that it is profitable for the attacker, thus reducing other miners' profits by the zero-sum nature of Bitcoin mining.
Both attacks have low hash power requirements to deploy, but their success rate improves substantially with increased hash power \cite{rosenfeld2011analysis}.
They can be covert over short periods, but over time produce visible signatures, making them overt over long periods.
\footnote{The attacks themselves reveal little information to the victim mining pool, but the change in mining patterns of the Chinese pools may be noticeable; for example, if a particular mining pool is receiving disproportionately fewer full blocks compared to the number of discovered blocks and observed Chinese mining hash power dedicated to mining on the Bitcoin network is decreasing, an observer can infer that Chinese mining pools are dedicating resources to mining pool attacks.}

\section{Conclusion}
\label{sec:conclusion}
As the value and economic utility of Bitcoin have grown, so has the incentive to attack it. We singled out China for analysis because they are the most powerful potential adversary to Bitcoin, and we found that they have a variety of salient motives for attacking the system and a number of mature capabilities, both regulatory and technical, to carry out those attacks. As future work, we suggest an analysis of existing solutions to the specific threats China poses to Bitcoin and the identification and mitigation of gaps in those protections.


\newpage 
\clearpage
\appendix
\section*{Appendix 1: Charting Bitcoin's Trajectory in China}
\label{apndx:china}
In this appendix, we briefly explore the factors that led to Bitcoin's dramatic growth in China and outline the evolution of the exchange and mining sectors and relevant regulations.

A number of factors set the stage in 2013 for Bitcoin to achieve popularity in China.
First, two economic trends in the aughts -- growth in private wealth and favorable foreign exchange rates for the Yuan -- drove increased buying power and desire for investment assets among consumers.
Because access to investment assets was (and still is) tightly controlled in China's command economy, Bitcoin was attractive due to its lack of regulation and potential for significant profit~\cite{Wong17}.
Second, by this time online and mobile payment systems were far more popular in China than anywhere else in the world~\cite{gloudeman_bitcoins_2014}. Bitcoin bears many similarities to such systems, especially when used through a mobile wallet app, so Chinese consumers may have been less hesitant to adopt Bitcoin than others.
Finally, China has long enforced a centralized political ideology and policies of strong social control. Banks in China are overwhelmingly controlled by the state, anonymous communication online is banned, and service providers are mandated to enforce Chinese state censorship on their platforms. Bitcoin represented an ideology of decentralization and individual autonomy that stood in direct opposition to these ideas, and its potential to provide a means of transacting anonymously and free from censorship was radical and appealing.

\subsection*{Exchanges}
Chinese exchanges grew to dominate the total exchange volume of Bitcoin from 2013 to 2017 (see Figure~\ref{fig:chinaexchange}). By the end of 2013, Chinese exchanges handled over 50\% of all trades and the single exchange BTC China was the largest in the world by trade volume, processing nearly a third of all transactions~\cite{McMillan13}.

Permitting large amounts of CNY to be converted to Bitcoin would seem to conflict with the protectionist economic policies of China, which include preventing capital flight -- the movement of capital out of China~\cite{ju2016capital}.
Bitcoin is a popular vector for capital flight because of its pseudonymity and limited oversight~\cite{drug2017national,amineh2018china}.
These qualities make Bitcoin attractive for other forms of crime as well, particularly money laundering. According to the 2017 National Drug Threat Assessment by the US Drug Enforcement Agency, ``China has been an enduring hub for trade-based money laundering schemes" whereby Chinese-made goods are purchased in bulk in USD and then sold in Mexico and South America for local currency.
``Chinese manufacturers who want Bitcoin undoubtedly ease the money laundering process" because criminal organizations can transfer large amounts of BTC to Chinese companies without incurring the scrutiny that would be applied were the funds in USD~\cite{drug2017national}.

When China announced their first Bitcoin regulations in December 2013, they cited criminal activity as the cause along with speculative risk~\cite{Mullany13}.
The policy banned financial institutions from buying and selling Bitcoin or treating it as a currency in any way.
Bitcoin itself was not made illegal, and exchanges were allowed to continue facilitating trades as long as they obeyed anti-money laundering regulations such as identifying users, but by cutting off the ability to exchange bitcoins for fiat currency, China undermined the main business model for Bitcoin exchanges and eliminated much of the utility of Bitcoin~\cite{Hill14}.

Chinese Bitcoin exchanges quickly exploited loopholes in the new regulation and deployed workarounds.
\footnote{These included selling voucher codes that customers could trade offline and redeem on the exchange, deploying physical ATMs that could exchange cash for bitcoins, and using corporate or even personal bank accounts to process transactions~\cite{Mu14,Bischoff14}.
}
Confusion ensued as regulators curbed some of this activity, but in the end most exchanges simply closed their accounts with Chinese commercial banks and used alternative financial systems.
\footnote{Following this confusion, the congressional US-China Economic and Security Review Commission observed in an issue brief that ``the true attitude of China's regulators towards Bitcoin is characteristically ambiguous"~\cite{gloudeman_bitcoins_2014}.}
 After a few months of uncertainty, the Chinese exchange market stabilized and exchange activity continued its surge with little regulatory interference for the next two years with CNY exchange comprising 98\% of all Bitcoin exchange activity in December 2016.
 \footnote{This value seems anomalously high and led to allegations of inflation or manipulation. One explanation is the unusual fee structure of Chinese exchanges in which trades are free and withdrawal fees decrease as a user's trade volume increases, incentivizing spurious trades. Varying estimates place China's real market share at this time as between 50\% and 85\% -- still a dominant position~\cite{Woo17}.}

In early 2017, regulators issued a series of warnings to exchanges to stay compliant with the 2013 policy~\cite{Rizzo17Feb,Rizzo17Jan}.
This regulation was accompanied by a ban on initial coin offerings (ICOs), an increasingly popular fundraising vehicle in which investors receive stake in the form of cryptocurrency tokens. Justifying the ban, regulators called ICOs "disruptive to economic and financial stability"~\cite{Zhao17}.
In the coming months, exchanges made some more efforts to improve anti-money laundering practices and curb speculation by instituting trading fees, but in September 2017 officials ordered the exchanges to shut down.
Loopholes such as over-the-counter sales, peer-to-peer trading, and foreign listings were banned in early 2018 forcing exchanges to finally abandon the Chinese market and relocate~\cite{scott_china_2018}. As a result, Chinese exchanges now accounts for less than 1\% of the global market.

\subsection*{Mining}
From January 2015 to January 2018, Chinese pools grew from accounting for 42\% of the total Bitcoin network hash power to 77\% (see Figure~\ref{fig:hashpower}).
\footnote{Prior to this time, block attribution is too sparse to reliably estimate relative share. We attribute the following pools to Chinese managers: AntPool, Bixin, BTCC, BTC.com, BTC.TOP, BW.COM, DPOOL, F2Pool, Poolin, ViaBTC, and 58COIN.}
Miners can contribute to Chinese-managed pools from all over the world, so these figures do not represent the share of hash power physically located in China.
However, China does host more mining facilities than any other country (data is not available to measure the precise share)~\cite{Hileman17}.
These facilities are primarily located in remote areas with inexpensive electricity and cheap land, such as Sichuan province and Inner Mongolia.
These advantages allow Chinese miners to achieve greater profit margins than their competitors in other countries; a study in early 2018 found that one bitcoin could be mined in China at $\frac{2}{3}$ the electricity cost of the same operation in the U.S.~\cite{Elitefixtures18}.

Another advantage enjoyed by Chinese miners is proximity to world-leading chip manufacturing facilities. In particular, the Application-Specific Integrated Circuits (ASICs) used for mining are overwhelmingly produced in China, with the most prominent manufacturer, BitMain, claiming to have produced 70\% of the chips used globally in 2017~\cite{Peck17}.

The growth of Chinese mining was further fueled by tax incentives and energy and land discounts offered by provincial governments. However, in early 2018 local regulators were directed by the central bank to ensure that Bitcoin miners no longer received preferential treatment~\cite{Yujian18} and shortly thereafter to scale down Bitcoin mining by regulating power usage, land use, taxes, and environmental protection~\cite{ReutersPolice18}. The stated motivation for the regulation was to make more electricity available for distribution to underserved regions, but the near-simultaneity with heavy exchange regulation suggests a focused effort to reduce Bitcoin's overall popularity and usage in China. These efforts are ongoing, but as Figure~\ref{fig:hashpower} shows, Chinese-managed mining pools remain dominant.

\clearpage
\bibliographystyle{splncs04}
\bibliography{chinab}
\normalsize

\end{document}